\documentstyle[aps]{revtex}
\begin{document}
\begin{center}
{\Large \bf
 The quark potential model for vector mesons \\and \\their decay constants.}\\
\vspace{2cm}
O. M. Juraev \\
{\sl The Abdus Salam International Centre For Theoretical Physics} \\
{\it P.O.BOX 586 - 34100 Trieste, Italy} \\
{\sl Institute of Nuclear Physics of Uzbekistan Academy of Sciences}\\ 
{\it Ulugbek, 702132, Tashkent, Uzbekistan}\\
\vspace{1cm}
T.\,Z.\,Nasyrov, Kh.\,Ablakulov, and  B.\,N.\,Kuranov,  \\
{\sl Institute of Nuclear Physics of Uzbekistan Academy of Sciences}\\ 
{\it Ulugbek, 702132, Tashkent, Uzbekistan}\\
\end{center}
\begin{abstract}
The  relativistic quark potential model (QPM)  is developed for describing
vector mesons and their leptonic decays. The new representation  Salpeter 
equation for vector mesons is proposed. Studying the $\tau\to\rho\nu$ decay the
expression for $\rho$-meson leptonic decay constant $f_\rho$ is
obtained. It is shown that quark model describes on a satisfactory level both the
constant $f_\rho$ and the masses of other charged vector mesons. The
values for the leptonic decay constants of $K^{*},D^{*},D^{*}_s,B^{*},B^{*}_c$ mesons 
are predicted.
\end{abstract}
\vspace{1cm}PACS number(s): 11.10.St; 12.39.-x; 12.39.Ki; 13.20.-v.
\vspace{2cm}
\section{Introduction}

The basic idea of our quark potential model  is simple. First, we consider 
Schwinger-Dayson equation which described one-particle quark "inside" hadron 
\cite{qrev}. The solutions of Schwinger-Dayson (SDE) we used in case bound 
state of quark antiquark pair.  Second step, we  consider Salpeter equation 
sometimes called the Bethe-Salpeter equation, which describe
the spectra mass of mesons as bound state of quark antiquark.
Using  the spectra masses, wave functions which we have found and then
we will calculate the leptonic decay constant of mesons.

At present a reliable experimental information on leptonic decay constants
of heavy pseudoscalar and vector mesons does not exist. However, these
constants play an important role in phenomenological describing of heavy mesons
physics: the $B-B_0$-splitting, mass difference and life time, violation of
$CP$-parity and  rare decays.

In the last time the mass spectrum of these mesons and their leptonic decay
constants is mainly described in the framework of phenomenological models
based on the ideas of the quantum chromodynamics (QCD) \cite{a1}-\cite{a4}.
The relativistic quark potential model  \cite{aBRPM}, on the basis of
which there are two principles -- {\it the minimal quantization of gauge
fields} and {\it a choice of the quantization axis} \cite{amin}, also belongs
to this kind of models.

In papers \cite{acal,apal,apal2} within this model the description of
spectroscopy of pseudoscalar mesons including pion had been obtained on a
satisfactorily level with their experimental values. However, the estimates
for the leptonic decay constant of pion $f_\pi$ were considerably smaller
than its experimental value \cite{apal,apal2}. The mass and decay constant of pion
are calculated in the framework of the quark potential model  in paper \cite{JurTurk}.
Radial excitation of $\pi', K', D'$ mesons and
their decay of constant was calculate in \cite{Sar} communication.
In paper \cite{ayaf1} the value of constant $f_\pi$ also could been reproduced by means
of modification of the SDE in this model. 

Using decomposition of the vector meson wave function over its structure
components in papers \cite{aBRPM,acal} the Salpeter equation (SE) for vector
mesons had been obtained. Then it was considered that the vector meson is
a bound state consisting of quark and antiquark and its polarization vectors
$\epsilon_\mu^\lambda$ ($\lambda=1,2,3$ are the polarization indices) depend
on relative momentum $q$.

However, it is known that the vector particle with nonzero mass having
internal structure is described by three orthogonal spacelike vectors
$\epsilon_\mu^\lambda$, which depend on total momentum ${\cal P}^\mu$ and
obey to orthogonality condition ${\cal P}^\mu\epsilon_\mu^i=0$. Hence it
follows that ${\cal P}^\mu$ plays role of the vector giving some
distinguished direction to which the vector fields are referred \cite{ashvin}.

In the present paper we proceed from that $\epsilon_\mu^\lambda$ are the
functions of ${\cal P}^\mu$ and propose new representation SE which
 describes mass spectrum
and wave functions of vector mesons. The $\tau\to\rho\nu$ decay is considered
and the expression for leptonic decay constant of $\rho$ meson $f_\rho$
is given. Choosing the interquark interaction potential as sum of the
oscillator and Coulomb type potentials and using solutions of the SDE and SE
proposed we obtain both constant $f_\rho$ and masses of other charged vector
mesons consisting of "up" and "down" quarks with different flavours on
qualitative level. The values for their leptonic decay constants are
predicted as well.

This paper is organized as follows. In Section II we obtain the new representation of 
SE for
vector mesons from the effective action of QCD for bilocal meson fields
and compare with its previous versions. In Section III we consider the
$\tau\to\rho\nu$ decay and representation for constant $f_\rho$ which can be
generalized also for the leptonic decay constants of other charged vector
mesons too is received. In Section IV we present the results of calculation of
the masses and leptonic decay constants of charged vector mesons and give an
outlook.
\section{The Salpeter equation for vector mesons }
We proceed from the effective action of QCD for bilocal meson fields
${\cal M}(x,y)$ proposed in Refs. \cite{aBRPM,acal}
\begin{equation}
W_{eff}({\cal M})=N_c\left\{\frac{1}{2}({\cal M},{\cal K}^{-1}{\cal M})
-i\mbox{\bf Tr}\ln\left[-(i\rlap/\partial-m^0)+{\cal M}\right]\right\},
\label{f-action}
\end{equation}
where
$${\cal K}(x,y)={\cal K}(z,X)=\rlap/\eta \times \rlap/\eta V(z^\bot)\delta(z\eta)$$
$$z=x-y,\ \ X=\frac{x+y}{2},\ \ z^\|=\eta(z\eta),\ \
z^\bot=z-z^\|, \ \ \rlap/\eta=\gamma_\mu\eta^\mu,\ \
\eta^\mu=\frac{{\cal P}^\mu}{\sqrt{{\cal P}^2}}.$$
Here $N_c=3$ is the number of colors, $V(z^\bot)$ is the phenomenological
interquark interaction potential which is used for describing of the
spectroscopy of mesons, $m^0=\mbox{diag}(m_1^0,...,m_{n_f}^0)$ is the matrix
of current quark masses, $n_f$ is the quark flavour number. The symbol {\bf Tr}
denotes integration over the space-time coordinates and summation over the
flavour and Dirac spinor indices.

Expansion of action (\ref{f-action}) about the stationary solution ($\Sigma$)
in small fluctuations\\
 ${\cal M}'$ (${\cal M}'={\cal M}-\Sigma$) leads to
the action bilinear in these fluctuations \cite{aBRPM}
\begin{equation}
W^{(2)}=N_c\left[\frac{1}{2}({\cal M}',{\cal K}^{-1}{\cal M}')+
\frac{1}{2}\mbox{\bf Tr}(G_\Sigma {\cal M}')^2\right],
\label{f-spectr}
\end{equation}
describing the meson spectrum, and to the interaction of these mesons
\begin{equation}
W_{int}=iN_c\mbox{\bf Tr}{\sum\limits_{n=2}^{\infty}}'
\frac{(G_\Sigma{\cal M}')^2}{n},
\label{f-int}
\end{equation}
where $G_\Sigma$ is the quark Green function, and prime $'$ denotes absence
of the quadratic term over the bilocal fields.

In order to account interaction of the bilocal meson fields between leptons
we are to modify ${\cal M}(x,y)$ taking the weak interaction into account
\begin{equation}
{\cal M}(x,y)\to {\cal M}(x,y)+{\cal L}(x,y),
\label{f-billep}
\end{equation}
where
$${\cal L}(x,y)=\frac{G_F}{\sqrt{2}}K_{ij}\gamma^\mu(1+\gamma_5)
l(x)\gamma_\mu(1+\gamma_5)\nu_l(y)\delta(x-y)$$
is the local leptonic current.\\
Here $K_{ij}$ are the Cabibbo-Kobayashi-Maskawa mixing matrix elements,
$G_F$ is the Fermi constant, $l(x)=(e(x),\,\mu(x),\,\tau(x))$ and
$\nu_l(x)=(\nu_e(x),\,\nu_\mu(x),\,\nu_\tau(x))$
are the leptons and their neutrinos wave functions, respectively.

Taking expression (\ref{f-billep}) into account from action (\ref{f-int}) we
get
\begin{equation}
W_{int}=iN_c{\sum\limits_{n=2}^{\infty}}'\frac{1}{n}
\left[G_\Sigma({\cal M}+{\cal L})\right]^n.
\label{f-int1}
\end{equation}

In order to calculate the matrix elements of the processes with
presence of a vector meson as a $\bar qq$-bound state, which is
described by effective QCD action (\ref{f-int1}) we decompose
${\cal M}(x,y)$ over the creation and annihilation operators of
mesons, $a^{\pm}_i({\cal P})$, with eigenvalues, $\sqrt{{\cal P}^2}$
$=M$ and $\omega=(\vec{\cal P}^2+M^2)^{1/2}$ ($M$ is the meson mass)
$$
{\cal M}(x,y)={\cal M}(z|X)=\sum\limits_{\lambda=1}^3
\int \frac{d^3\vec{\cal P}}{(2\pi)^{3/2}\sqrt{2\omega}}
\int \frac{d^4q}{(2\pi)^4}\exp(iqz)
[
\exp(i{\cal P}X)\Gamma^\lambda(q|{\cal P}) a_\lambda^{+}({\cal P})
$$
\begin{equation}
+\exp(-i{\cal P}X)\bar\Gamma^\lambda(q|{\cal P}) a_\lambda^{-}({\cal P})],
\label{f-rojd}
\end{equation}
where $\Gamma^\lambda(q|{\cal P})$ and $\bar\Gamma^\lambda(q|{\cal P})$
are the vector meson vertex functions.

In expression (\ref{f-rojd}) operators $a^{+}_i({\cal P})$ and $a_i^{-}({\cal P})$
satisfy to commutative relations
$$[a_\lambda^{-}({\cal P}),a_{\lambda'}^{+}({\cal P}')]_{-}=
\delta_{\lambda\lambda'}
\delta(\vec{\cal P}-\vec{\cal P}'),\qquad
[a_\lambda^{\pm}({\cal P}),a_{\lambda'}^{\pm}({\cal P}')]_{-}=0.$$

Variation of action (\ref{f-spectr}) about of ${\cal M}'$ taking into account
expression (\ref{f-rojd}) leads to the Bethe-Salpeter equation
\begin{equation}
\Gamma_{(a,b)}^\lambda(p|{\cal P})=-i\int\frac{d^4q}{(2\pi)^4}V(p^\bot-q^\bot)
\rlap/\eta G_{\Sigma_a}\left(q+\frac{{\cal P}}{2}\right)
\Gamma_{(a,b)}^\lambda(q|{\cal P}) G_{\Sigma_b}
\left(q-\frac{{\cal P}}{2}\right)\rlap/\eta,
\label{f-ubs}
\end{equation}
where $a,b$ are the quark and antiquark flavours.

Integrating equation (\ref{f-ubs}) over the longitudinal momentum
$q^\|=\eta_\mu(q\eta)$ and introducing the "dressed" vector meson wave
function $\Psi_{(a,b)}^\lambda(q^\bot|{\cal P})$
\begin{equation}
\Psi_{(a,b)}^\lambda(q^\bot|{\cal P})=
S_a(q^\bot)\Psi_0^\lambda(q^\bot|{\cal P})S_b(q^\bot)=
i\int\frac{dq^\|}{2\pi}G_{\Sigma_a}
\left(q+\frac{{\cal P}}{2}\right)\Gamma_{(a,b)}^\lambda(q^\bot|{\cal P})
G_{\Sigma_b}\left(q-\frac{{\cal P}}{2}\right),
\label{f-vol}
\end{equation}
where
\begin{equation}
S_{a,b}(q^\bot)=\exp\left[\frac{\rlap/ q^{\bot}}{|q^\bot|}
\vartheta_{a,b}(q^\bot)\right]
\label{f-folvaut}
\end{equation}
is the Foldy-Wouthuysen type transformation matrix \cite{aBRPM},
\begin{equation}
\Psi_0^\lambda(q^\bot|{\cal P})=\rlap/\epsilon^\lambda
({\cal P})(N_1(q^\bot)-\rlap/\eta
N_2(q^\bot))
\label{f-razloj}
\end{equation}
is the  "undressed" vector meson wave function,
we obtain the SE for vector mesons in the following form
$$
MN_{2\choose 1}(p^\bot) =
E_t(p^\bot) N_{1\choose 2}(p^\bot)
-\frac{1}{3}
\int\frac{d^3q^\bot}{(2\pi)^3}V(p^\bot-q^\bot)
[c^{\mp}_pc^{\mp}_q + c^{+}_p c^{-}_q + c^{-}_p c^{+}_q$$
%\right. 
\begin{equation}
% \left.
- \xi (2s^{\mp}_p s^{\mp}_q+ s^{\pm}_p s^{\pm}_q)+
\xi^2 (c^{-}_pc^{-}_q + c^{+}_p c^{-}_q + c^{-}_p c^{+}_q + c^{+}_pc^{+}_q)]
N_{1\choose 2}(q^\bot),
\label{f-salp}
\end{equation}
where
$$c^{\pm}_p=\cos[\vartheta_a(p^\bot)\pm \vartheta_b(p^\bot)],\quad
s^{\pm}_p=\sin[\vartheta_a(p^\bot)\pm \vartheta_b(p^\bot)],\quad
\xi=\frac{p^\bot q^\bot}{|p^\bot||q^\bot|},$$
$$ E_t(p^\bot)=E_a(p^\bot)+E_b(p^\bot).$$
Here $\vartheta_{a,b}(p^\bot)$ and $E_{a,b}(p^\bot)$ are the single-particle
phase functions and energies of quark and antiquark inside the meson,
which are found by solving the SDE \cite{qrev}
\begin{equation}
E_{a,b}(p^\bot)\cos2\vartheta_{a,b}(p^\bot)=m^0+\displaystyle{\frac{1}{2}\int}
\frac{d^3q^\bot}{(2\pi)^3}
V(p^\bot - q^\bot) \cos2\vartheta_{a,b}(q^\bot),
\label{f-usd1}
\end{equation}
\begin{equation}
E_{a,b}(p^\bot)\sin2\vartheta_{a,b}(p^\bot)=|p^\bot|+
\displaystyle{\frac{1}{2}\int}\frac{d^3q^\bot}{(2\pi)^3}
V(p^\bot - q^\bot) \xi \sin2\vartheta_{a,b}(q^\bot). 
\label{f-usd2}
\end{equation}

In expression (\ref{f-razloj}) $N_1(q^\bot)$ and $N_2(q^\bot)$ are the structure
formfactors of the vector meson, $\epsilon^\mu_i({\cal P})$ are its
polarization vectors, which satisfy the following conditions, respectively:
%\begin{eqnarray*}
$$\frac{4N_c}{M}\int \frac{d^3q^\bot}{(2\pi)^3}N_1(q^\bot)
N_2(q^\bot) =1,$$ \\
%\label{f-Norm}\end{eqnarray*}
$$\sum\limits_{\lambda=1}^3
\epsilon_\alpha^\lambda({\cal P})\epsilon_\beta^\lambda({\cal P})=-
\left(g_{\alpha\beta}-\frac{{\cal P}_\alpha
{\cal P}_\beta}{M^2}\right).$$

It is to be noted that unlike the works \cite{aBRPM,acal}, where $N_1$
and $N_2$ are the vector quantities, in our scheme they are divided to
the polarization vectors matrix and structure scalar formfactors (see, expression
(\ref{f-razloj})), which are decomposed over the spherical functions. This
circumstance simplifies to find the mass spectrum and wave functions
of vector mesons, and leads to the only leptonic decay constant.
\section{$\tau\to\rho\nu$ decay and leptonic decay constants of vector mesons}
Now we consider $\tau\to\rho\nu$ decay, studying of which gives an
information about of hadronization of the intermediate charged weak
vector $W$ boson to the vector meson, that allows to calculate the
constant $f_\rho$.

The matrix element of the decay is
\begin{equation}
<\rho\nu|W_{int}|\tau>=-N_c\mbox{\bf Tr}\int d^4x_1 d^4x_2 d^4y_1 d^4y_2
<\nu|{\cal L}(x_1,y_2)|\tau><\rho|G_\Sigma(x_1-x_2){\cal M}(x_2,y_1)
G_\Sigma(y_1-y_2)|0>,
\label{f-matr}
\end{equation}
where
\begin{eqnarray*}
G_\Sigma(x-y)=\left(\begin{array}{cc}
G_{\Sigma_u}(x-y)&0\\0&G_{\Sigma_d}(x-y)
\end{array}\right)
%\label{f-grin1}
\end{eqnarray*}
is the Green functions matrix of $u$ and $d$ quarks;
\begin{eqnarray*}
{\cal M}(x,y)=\left(\begin{array}{cc}
\displaystyle{\frac{{\cal M}_{\rho^0}(x,y)}{\sqrt{2}}}&
{\cal M}_{\rho^{+}}(x,y)\\{\cal M}_{\rho^{-}}(x,y)&
-\displaystyle{\frac{{\cal M}_{\rho^0}(x,y)}{\sqrt{2}}}
\end{array}\right)
%\label{f-biloc}
\end{eqnarray*}
is the matrix of bilocal meson fields;
$${\cal L}(x,y)=\tau^{-}\gamma_\alpha(1-\gamma_5)\delta^4(x-y){\cal L}^\alpha(x),$$
%\label{f-alocal}\end{eqnarray*}
is the local leptonic current.\\ Here
%\begin{eqnarray*}
$${\cal L}^\alpha(x)=\frac{G_F\cos\theta_c}{\sqrt{2}}
\bar\psi_\nu(x)\gamma^\alpha(1-\gamma_5)\psi_\tau(x),$$
%\label{f-local}\end{eqnarray*}
$\tau^{-}=(\tau_1-\tau_2)/2$, $\vec\tau$ is the Pauli matrices,
$\psi_\tau(x)$ and $\psi_\nu$ are the $\tau$ and its neutrino
wave functions, respectively.

From expression (\ref{f-matr}) using relations (\ref{f-int1}) and (\ref{f-rojd})
for the matrix element we obtain
\begin{equation}
<\rho\nu|W_{int}^\lambda|\tau>=\frac{(2\pi)^4\delta(k_\tau-{\cal P}-k_\nu)}
{[(2\pi)^9\cdot2\omega_\tau\cdot 2\omega\cdot 2\omega_\nu]^{1/2}}
{\cal L}^\alpha(k_\tau,k_\nu){\cal H}^\lambda_\alpha({\cal P}),
\label{f-halfa}
\end{equation}
where ${\cal L}^\alpha(k_\tau,k_\nu)$ is the leptonic part of the matrix
element \cite{aokun}, and ${\cal H}^\lambda_\alpha({\cal P})$ is its hadronic
part which within QPM has the following form
\begin{equation}
{\cal H}^\lambda_\alpha({\cal P})=-i N_c\int\frac{d^4q}{(2\pi)^4}\mbox{tr}
\left[G_{\Sigma_u}\left(q+\frac{{\cal P}}{2}\right)
\Gamma_{(u,d)}^\lambda(q|{\cal P}) G_{\Sigma_d}
\left(q-\frac{{\cal P}}{2}\right)\gamma_\alpha(1-\gamma_5)\right].
\label{f-hadrn}
\end{equation}

After integration of hadronic part (\ref{f-hadrn}) over $q^\|$ taking into
account expression (\ref{f-vol}) we have
\begin{eqnarray}
{\cal H}^\lambda_\alpha({\cal P})=N_c\int\frac{d^3q^\bot}{(2\pi)^3}
\mbox{tr}\left[ S_u(q^\bot)\Psi^\lambda_0(q^\bot|{\cal P})S_d(q^\bot)
\gamma_\alpha(1-\gamma_5) \right].
\label{f-hadron}
\end{eqnarray}

Hence taking into account expressions (\ref{f-razloj}) and (\ref{f-folvaut}) we
find
\begin{equation}
{\cal H}^\lambda_\alpha({\cal P})=f_\rho\epsilon_\alpha^\lambda({\cal P}),
\label{f-hadkon}
\end{equation}
where
\begin{equation}
f_\rho=\frac{4N_c}{3}
\int\frac{d^3q^\bot}{(2\pi)^3}N_1(q^\bot)
 \left\{2\cos\left[\vartheta_u(q^\bot)-
\vartheta_d(q^\bot)\right]+\cos\left[\vartheta_u(q^\bot)+
\vartheta_d(q^\bot)\right]\right\}
\label{f-const}
\end{equation}
is the leptonic decay constant of $\rho$ meson.

It is seen from expression (\ref{f-const}) that the constant $f_\rho$ is expressed
in terms of solution of the SDE for the single-particle phase
functions of $u$ and $d$ quarks inside $\rho$ meson and solution of the SE
suggested for the wave function itself.

It should be noted that representation (\ref{f-const}) can be used also
for defining leptonic decay constants of other charged vector mesons
consisting of "up" and "down" quarks too.
\section{Results of numerical calculations and discussion}
Proceeding from the fact that the oscillator potential leads to spontaneous
breakdown of the chiral symmetry, which describes on qualitative level
the large difference of masses between $\pi$ and $\rho$ mesons \cite{a1}, and
Coulomb type -- to the asymptotically freedom \cite{asymp}, we use the
following interquark interaction potential
\begin{equation}
V(r)= \frac{4}{3}\left(V_0r^2-\frac{\alpha_s}{r}\right), \label{f-potinr}
\end{equation}
where $V_0$ and $\alpha_s$ are the parameters of their potentials.

As a rule SDE (\ref{f-usd1}),(\ref{f-usd2}) and SE (\ref{f-salp}) with such kind of potentials
are solved using of numerical methods. In order to remove ultraviolet
divergences in the SDE, arising because of Coulomb part of the potential we
use the standard renormalization scheme, proposed in Refs. \cite{a4}.

In paper \cite{a1} mass spectrum of mesons had been calculated by solving
the SDE and SE with pure oscillator potential ($\alpha_s=0$). Then it was
considered that the current quark mass equals to zero ($m^0=0$) and pion is
the pseudo Goldstone particle with zero mass. Fitting only parameter $V_0$
by the $\rho$ meson mass for it was obtained $(4V_0/3)^{1/3}=289$ {\it MeV},
which allows to describe on a satisfactory level the mass spectrum of light
mesons.

However, the spectroscopy of heavy mesons and also leptonic decay constants of
pseudoscalar mesons could not been reproduced with oscillator potential.
Taking into account availability of the current masses $u$ and $d$ quarks,
values of which were defined fitting by pion mass ($m^0_u=m^0_d=2$ {\it MeV}),
does not resolve this problem either \cite{apal}.

Complication of the potential by means of adding to potential of the Coulomb
type term ($\alpha_s\not= 0$), which is due to one gluon exchange between
quarks and also taking into account other quark ($s$,\, $c$ and $b$) flavours,
of course, leads to redefinition of values of input parameters $V_0$ and $m^0$.
After fitting them by masses of $\pi$,\, $\rho$,\, $K$,\, $D$ and $B$ mesons
we receive
$$(4V_0/3)^{1/3}=299\ \ MeV,\qquad
\alpha_s=0.2\qquad m^0_{u,d}=2.3\ \ MeV, $$
$$m^0_s=68\ \ MeV,\qquad m^0_c=1273\ \ MeV,\qquad m^0_b=4720\ \ MeV.$$

Solving equation (\ref{f-salp}) with potential (\ref{f-potinr}) numerically by
Continuous Analogy of Newton Methods \cite{aNewton2} using of shown solutions 
of the SDE from expression
(\ref{f-const}) we obtain the following value for constant $f_\rho$
$$f_\rho=1.03\cdot 10^5\,\,MeV^2, $$
which is closer to available experimental data
\cite{aeksp1} $$f_\rho^{exp}\approx 1.57\pm 0.09\cdot 10^5\,\,MeV^2. $$

As for the values of masses of other charged vector mesons and their leptonic
decay constants calculated within QPM with potential (\ref{f-potinr})
they are listed in Table 1.  \vspace{5mm} \\
Table 1. Values of masses of vector mesons and their
leptonic decay constants ($(4V_0/3)^{1/3}=299$ {\it MeV},\, $\alpha_s=0.2$,\,
$m^0_{u,d}=2.3$ {\it MeV},\, $m^0_s=68$ {\it MeV},\, $m^0_c=1273$ {\it MeV},\,
$m^0_b=4720$ {\it MeV}). \vspace{2mm} \\
\begin{center}
\begin{tabular}{|c|c|c|c|c|}\hline
\multicolumn{1}{|c|}{Vector}&\multicolumn{2}{c|}{Masses, ({\it MeV})}&
\multicolumn{2}{c|}{Decay Constants, $\times 10^5\ \ MeV^2$}\\ \cline{2-5}
meson &Model&Experiment \cite{aeksp}&Model&Experiment \\ \hline
$\rho$&\underline{770}&770&1.03&$1.12\pm 0.09$ \cite{aeksp1}\\ \hline
$K^*$&807&892&1.17&\\ \hline $D^*$&1880&2010&3.21&\\ \hline
$D_s^*$&1894&2110&3.51&\\ \hline $B^*$&5281&5325&6.61&\\
\hline $B_c^*$&6203&&15.28&\\ \hline
\end{tabular}\vspace{5mm}
\end{center}
It is seen from the Table 1 that QPM with potential (\ref{f-potinr})
can reproduce in satisfactory level the masses of charged vector mesons.

Thus, we would like to note that calculation of the leptonic decay constants
both pseudoscalar and vector mesons is important for extraction from
experimental data of corresponding Cabibbo-Kobayashi-Maskawa matrix elements,
 which are input
parameters of standard model of interaction of elementary particles.
We have the only charged lepton and its neutrino in the end of leptonic decays
of mesons and then effect of strong interactions in the initial state
is parametrized to the only constant. Consequently, these constants within
QPM with SDE for the quark phase function and SE for the bound state of
quarks describe hadronization of the $W$ boson to the charged bilocal vector
meson, and vice versa, annihilation of the vector meson to charged weak boson.

\section{ACKNOWLEDGMENTS}

One of the authors (OMJ) would like to thank S.Randjbar-Daemi for invitation and 
 hospitality in the Abdus Salam International Centre for Theoretical Physics. 
Authors thank to N.A. Sarikov for stimulating discussions of the obtained 
results and B.S. Yuldashev for support.

\end{document}